\documentstyle[preprint,floats,aps,graphicx]{revtex}

\def\sla{/\!\!\!}
\newcommand{\Over}[2]{%
\raisebox{0.4ex}{\textrm{\footnotesize $#1$}}/
\raisebox{-0.4ex}{\textrm{\footnotesize $#2$}}}

\begin{document}

\preprint{\vbox{ \hbox{\em \today}   \hbox{hep-ph/yymmxxx} }}

\title{Exclusive Semileptonic Decays of B Mesons \\
to Orbitally and Radial Excited D}

\author{Massimo Di~Pierro and Adam~K. Leibovich}

\address{Theory Group, Fermi National Accelerator Laboratory,\\ 
         Batavia, IL 60510, USA}

\maketitle

\begin{abstract}

In this paper we compute, within in the context of a relativistic quark 
model, the Isgur-Wise functions for exclusive semileptonic 
$\bar B \to X_c$ decays, where $X_c$ is any charmed mesons with total spin 
$J=0,1,2$ or one of their first excited states. The relevant matrix elements 
are computed by a direct numerical integration, in coordinate space, of the 
convolution of the wave function of the $B$ meson at rest and the wave 
function of the $X_c$ meson, boosted according with its recoil factor.
Our results are compared with other predictions found in 
the existing literature.

\end{abstract}

\newpage

\section{Introduction}

There has been much progress in recent years in model-independent
calculations of heavy meson decays.  By using Heavy Quark Effective
Theory (HQET) \cite{hqet} and Lattice Gauge Theory, we can now make
some definite predictions for certain processes, in a limited
kinematic range.  Knowledge of these decays are extremely important
for particle physics, not just in their own right, but in measuring
fundamental parameters such as $V_{cb}$ and $\sin(2\beta)$.

Unfortunately, at present time, it is not always possible to use 
these techniques for all kinematic situations. 
One example are the Isgur-Wise (IW) functions, which relate all the
different form factors for heavy-to-heavy decays to a single function,
at leading order in $\Lambda/m$.  These functions can,
in principle, be determined by lattice computations but, 
because of limited computing resources, they have only been computed
for decays into the ground state and the error is still sizable.
For the moment HQET can be used to produce model independent 
results at zero recoil but, away from this
point, the IW functions are unknown.  
Therefore, for the time being, we must rely on models.

We present here a study of $\bar B$ decaying into
excited $D$s and a determination of the corresponding IW functions, 
following the work of Ref.~\cite{faustov}. Our study is based on the 
quark model proposed in Ref.~\cite{dipierro}, where a Dirac equation 
was used to describe the light quark in the potential of the heavy 
quark and determine masses and wave functions of excited mesons.
We use these wavefunctions to calculate the leading order 
IW functions for $\bar B\to X_c$ decays, where $X_c$ is a spin 0 -- 2 charmed 
meson or its first radially excited state.  
The IW functions are computed explicitly by
a three dimensional numerical integration of the relevant matrix
elements expressed in terms of the wave functions derived in
Ref.~\cite{dipierro}.

The paper is organized as follows.  In Section II we set up the
formalism and in Section III we discuss the actual calculation of the
IW functions.  In Section IV we discuss the results and compare with
the literature.  Finally in Section V we conclude.

\section{The Quark Model}

Due to heavy quark symmetry, we know a great deal about the spectroscopy of
mesons containing a heavy quark $Q$.  At leading order in
$\Lambda/m_{Q}$, the spin and parity of the light degrees of freedom,
$s_l^{\pi_l}$, are good quantum numbers.  Thus the particles appear in
multiplets labeled by both the total spin $J$ and $s_l^{\pi_l}$.  In
this limit, the mesons come in degenerate doublets with total spin
$J_\pm = s_l \pm s_Q$, where $s_Q = 1/2$ is the heavy quark spin,
a conserved quantum number at leading order in HQET.
In Table \ref{particles} we list the low spin charmed mesons considered 
in this paper, their masses and corresponding quantum numbers in HQET. 
In those cases where the experimental mass is not known we report, 
in brackets, the mass predicted in Ref.~\cite{dipierro}. For the mass of the
decaying $\bar B$ meson we adopt the PDG value $m_B=5.279$ GeV.

{\tighten
\begin{table}[t]
\begin{tabular}{cccccccc}
& $j^P \equiv s_l^{\pi_l}$ & Particles       &  $J^P$  & $m$ (GeV) 
& $m'$ (GeV) \\ \hline 
& $\frac12^-$  & $D$       &  $0^-$ & $1.865$  & $(2.589)$\\
&   &  $D^*$       &  $1^-$ & $2.007$  & $(2.692)$\\
& $\frac12^+$  & $D_0^*$ &  $0^+$ & $(2.377)$ & $(2.949)$ \\
&   &  $D_1^*$ &  $1^+$ & $(2.490)$ & $(3.045)$ \\
& $\frac32^+$  & $D_1$   &  $1^+$ & $2.422$ & $(2.995)$ \\
&   & $D_2^*$   &   $2^+$ & $2.459$ & $(3.035)$ \\
& $\frac32^-$  & $D_1^{**}$   &  $1^-$ & $(2.795)$ & $(3.420)$ \\
&   & $D_2^{**}$   &  $2^-$ & $(2.833)$ & $(3.459)$ \\
\end{tabular} \vspace{6pt}
\caption{Charmed meson spin multiplets ($q=u,d$).  The masses are
experimental when possible, otherwise they are calculated from the
model.  The primed masses are for the first radial excited states.
The masses in parenthesis are predictions of the in the model.} 
\label{particles}
\end{table}
}

The model of Ref.~\cite{dipierro} is a relativistic quark model
characterized by a spin-dependent plus a spin-independent
potential. The general form for these potentials has been derived by
general arguments such as confinement and asymptotic freedom.  The
parameters that appear in the potential have been fixed by fitting the
experimental spectrum of $D$, $D_s$, $B$ and $B_s$ mesons with the
predicted spectrum, including $1/m_b$ corrections~\cite{roberts} 
and mixing effects.

The most general heavy-light meson (in the ${D, D_s, B, B_s}$ family),
$H$, is a bound state of a light quark ($q$) and a heavy quark
($Q$). The heavy quark is treated as a static source of chromoelectric
field and the only quantum number associated with it is its spin. The
light quark is treated relativistically and its state is described by
the wavefunction $\psi_{n,\ell ,j,m}(r,\theta ,\varphi)$.  We
introduce the following quantum numbers:

\begin{itemize}{\tighten
\item  $n$, the number associated with the radial excitations

\item  $\ell$, the orbital angular momentum

\item  $j$, a short hand notation for $s_\ell$, the total angular 
momentum of the light quark

\item  $m$, the component of $j$ along the $\widehat{z}$ axis

\item  $J$, the total angular momentum of the system

\item  $M$, the component of $J$ along the $\widehat{z}$ axis

\item  $S$, the spin of the heavy quark along the $\widehat{z}$ axis 
}
\end{itemize}
The parameters of the model are the masses of the light quarks ($m_q$
for $q = u, d {\rm ~or~} s$), the masses of the heavy quarks ($m_Q$
for $Q = c {\rm ~or~}b$) and the chromoelectric potential of the heavy
quark ($V(r)$).

The total wavefunction of the system can be decomposed as follows
\begin{equation}
\Psi _{n,\ell ,j,J,M}(r,\theta ,\varphi )=\sum_{S\in \left\{ -\frac{1}{2},+
\frac{1}{2}\right\} }C_{j,m;\frac{1}{2},S}^{J,M}\psi_{n,\ell ,j,m}(r,\theta
,\varphi) \otimes \xi_{S},  \label{notation0}
\end{equation}
where $C_{j,m;\frac{1}{2},S}^{J,M}$ are the usual Clebsh-Gordan
coefficients and $\xi _{S}$ is a two component spinor representing the
heavy quark.
In Eq.~(\ref{notation0}), the four spin components of the light quark
wavefunction, is parametrized as follows:
\begin{equation}
\Psi _{n,\ell ,j,J,M}(r,\theta ,\varphi ) = 
\hspace{-0.1cm}\sum_{S\in \left\{-\frac 12,+\frac 12\right\} }
C_{j,m;\frac 12,S}^{J,M}\left( 
\begin{array}{l}
if_{n,\ell ,j}^0(r)k_{\ell ,j,m}^{+}\,\,Y_{m-\frac 12}^\ell (\theta ,\varphi )
\\ 
if_{n,\ell ,j}^0(r)k_{\ell ,j,m}^{-}\,\,Y_{m+\frac 12}^\ell (\theta ,\varphi )
\\ 
\,f_{n,\ell ,j}^1(r)k_{2j-\ell ,j,m}^{+}
  Y_{m-\frac 12}^{2j-\ell }(\theta
,\varphi ) 
\\ 
\,f_{n,\ell ,j}^1(r)k_{2j-\ell ,j,m}^{-}
  Y_{m+\frac 12}^{2j-\ell }(\theta,\varphi)
\end{array}
\right) \otimes \xi_S .  \label{notation}
\end{equation}
Here $Y_m^\ell (\theta ,\varphi)$ are spherical harmonics that encode
the angular dependence while $f_{n,\ell,j}^0(r)$, $f_{n,\ell,j}^1(r)$
are real functions that encode that radial dependence.
$k_{\ell,j,m}^{+}$ and $k_{\ell ,j,m}^{-}$ are fixed, up to an
overall phase, by imposing a normalization condition. Our choice of
the phase is such that

\begin{equation}
k_{\ell ,j,m}^{\pm }=\left\{ 
\begin{array}{ll}
+\sqrt{\frac{\ell \pm m+\frac 12}{2\ell +1}} 
  & \text{for }j=\ell +\frac12 
\\ 
\pm \sqrt{\frac{\ell \mp m+\frac 12}{2\ell +1}} 
  & \text{for }j=\ell -\frac12
\end{array}
\right. . \label{kappas}
\end{equation}

The Hamiltonian of the most general heavy-light system, at leading order
in $1/m_b$ reads
\begin{equation}
{\cal H}^{(0)}=\gamma ^0(-i \sla \partial +m_q)+V(r),  \label{dirac2}
\end{equation}
and the rotational-invariant potential is the sum of a constant factor
($M_Q$), a scalar part ($V_s$) and (the zeroth component of) a vector
part ($V_v$)
\begin{equation}
V(r)=M_Q+\gamma ^0V_s(r)+V_v(r),  \label{potential}
\end{equation}
where
\begin{eqnarray}
V_v(r)&=&-\frac43\frac{\alpha _s}r\text{erf}(\lambda r), \label{potential1} 
\\
V_s(r)&=&br+c .  \label{potential2}
\end{eqnarray}
The role of the erf() in the potential is that of regularizing the
ultraviolet divergence in the $1/m_b$ corrections to the spectrum.
The parameters of the model, determined in Ref.~\cite{dipierro}, are
shown in Table \ref{parameters}. 

\begin{table}[t]
\begin{tabular}{ccccccccc}
$\alpha _s$ & $\lambda $ & $b$ & $m_u$ & $m_s$ & $m_c$ & $M_c$ & $m_b$ & $M_b$ 
\\
\hline
0.339 & 
2.823 &
0.257 & 
0.071 &
0.216 &
1.511 & 
1.292 & 
4.655 & 
4.685  
\end{tabular} \vspace{6pt}
\caption{Parameters for the model.  The masses and $\lambda$ are all
measured in GeV.  The parameter $b$ is measured in GeV$^2$.}
\label{parameters}
\end{table}

\section{Calculation}

The hadronic part of the semileptonic exclusive decay
$\bar B \rightarrow D+\ell+\bar\nu$ (for the most general 
excited D in the final state) is encoded in
a matrix element of the form
$\langle D(n',\ell',j',J',M') | \Gamma |\bar B(n,\ell,j,J,M)\rangle $
where  $\hat \Gamma = \bar u_c(0) \Gamma u_b(0)$.
For the decays of interest $\Gamma=\gamma^\mu$ or $\gamma^\mu\gamma^5$
but for the purpose 
of this paper we consider the most general $\Gamma$ structure. In fact,
thanks to the HQET, all matrix elements that differ only for the spin 
structure can be related to the same IW function.
The heavy-light states are normalized according with the usual 
non-relativistic convention.

A general formalism for the computation of matrix elements between 
states represented by wave functions was derived in 
Ref.~\cite{faustov}. In that paper the problem of defining equal 
time wave functions is discussed and the matrix elements are written
as a integral in momentum space of the Fourier transformed wave 
functions.
The general problem is greatly simplified in the specific case of 
heavy-light systems since it is always possible to shift a meson 
in time by changing the phase of the heavy quark.
Hence, in this paper, we find more intuitive to 
express our matrix elements as integrals in coordinate space which,
in the most general case of interest, look like:

\begin{equation}
\langle D(n',\ell',j',J',M') | 
\Gamma |\bar B(n,\ell,j,J,M)\rangle = \int 
\Psi_{n', \ell',j',J',M'}^{\mathbf p\dag}({\mathbf x}) 
 \hat \Gamma \Psi_{n,\ell,j,J,M}({\mathbf x})\,
\text{d}^3{\mathbf x},
\label{matrixelem1}
\end{equation}
where  $\Psi^{\mathbf p}$ is the wavefunction $\Psi$ boosted according
with the recoil momentum ${\mathbf p}$:
\begin{equation}
\Psi_{n,\ell ,j,J,M}^{\mathbf p}({\mathbf x}) = 
\sum_{S\in \left\{ -\frac{1}{2},+\frac{1}{2}\right\} }
C_{j,m;\frac{1}{2},S}^{J,M}\psi^{\mathbf p} _{n,\ell ,j,m}(
{\mathbf x}) \otimes \xi^{\mathbf p} _{S},  \label{notation1}
\end{equation}
and 
\begin{eqnarray}
\psi^{\mathbf p}({\mathbf x}) 
&=& S(\Lambda_{\mathbf p})\psi(\Lambda_{\mathbf p} ^{-1} \mathbf x), \\
\xi^{\mathbf p} _{S} &=& S(\Lambda_{\mathbf p})\xi_{S}.
\end{eqnarray}
where $\Lambda_{\mathbf p}$ is a boost in direction ${\mathbf p}$.

Note for any state only that component of the spin parallel to the
direction of motion ${\bf p}$ is a good conserved quantum number, the
helicity. Therefore in Eq.~(\ref{matrixelem1}) we chose $m'$, $M'$ and
$S$ to be the components of the light angular momentum, the total
angular momentum and the heavy quark spin respectively, parallel to
the direction of the boost $\Lambda_{\bf p}$. We checked that, with
this definition, the result for the matrix element is independent on
the direction of ${\mathbf p}$.  In our analysis we ignore mixing in
the wavefunctions and other $O(m_b^{-1})$
corrections.\footnote{Ref.~\cite{amundson} suggests that $1/m_b$
corrections to the matrix element corresponding to $\bar B \rightarrow
D^{(*)}$ are smaller than expected by dimensional analysis. We do not
know if this is also the case for the other matrix elements of
relevance in this paper.}  The 3D integrals are evaluated numerically
using the Vegas Monte Carlo algorithm.

In order to compare matrix elements we calculate with the model to
the corresponding IW functions, we need to calculate the matrix
elements in HQET.  This can be done by using the trace formalism
\cite{falk,trace,falk2}.  As an example, consider the $s_l^{\pi_l} =
{\frac12}^-$ doublet.  The fields $P_v$ and $P_v^{\ast\mu}$ that
destroy the members of this doublet with four-velocity $v$ are grouped
together in the $4\times4$ matrix
\begin{equation}
H_v^{-} = \frac{1+\sla v}2\left[P_v^{\ast\mu}\gamma_\mu -
P_v\gamma_5\right].
\label{hminus}
\end{equation}
The matrix $H_v^-$ satisfies the relations $\sla v H_v^- = H_v^- = -
H_v^- \sla v$.  To leading order in $\Lambda_{\rm QCD}/m_{b,c}$ and
$\alpha_s$, the matrix element between $\bar B^{(*)}$ and $D^{(*)}$ mesons
are

\begin{equation}
\langle D^{(*)} | \bar c \Gamma b | \bar B^{(*)} \rangle 
 = \xi(w) Tr[\bar H_{v'}^{-,c} \Gamma H_v^{-,b}].
\label{matrix1}
\end{equation}
Here $\xi(w\equiv v\cdot v')$ is the (dimensionless) IW function for 
$\bar B$ decaying to
the $D^{(*)}$ multiplet.  Matrix elements with any $\Gamma$ can now
easily be calculated and are related by heavy quark symmetry to
$\xi(w)$.

For other multiplets, different $4\times4$ matrices are used.  The
general form for arbitrary spin was developed in Ref.~\cite{falk2}.
Below are the other matrices which are necessary to calculate relations used
in this paper:
\begin{eqnarray}
H_v^+ &=& \frac{1+\sla v}2\left[P^{\ast\ \mu}_v\gamma_\mu \gamma^5 - P_v
	\right]\, , \label{hplus}\\
F_v^{+,\mu} &=& \frac{1+\sla v}2\, \bigg\{ P_v^{*\mu\nu} \gamma_\nu 
  - \sqrt{\frac32}\, P_v^\nu \gamma_5 \bigg[ g^\mu_\nu - 
  \frac13 \gamma_\nu (\gamma^\mu-v^\mu) \bigg] \bigg\} \,,\\
F_v^{-,\mu} &=& \frac{1+\sla v}2\, \bigg\{ P_v^{*\mu\nu} \gamma_\nu \gamma_5
  - \sqrt{\frac32}\, P_v^\nu \bigg[ g^\mu_\nu - 
  \frac13 \gamma_\nu (\gamma^\mu+v^\mu) \bigg] \bigg\} \,,
\label{fpm}
\end{eqnarray}
where $H_v^+$ is for the $s_l^{\pi_l} = {\frac12}^+$ doublet, and
$F_v^{\pm,\mu}$ are for the $s_l^{\pi_l} = {\frac32}^\pm$ doublets.

Using the trace formalism, we can relate the matrix elements
calculated in the model, Eq.~(\ref{matrixelem1}), to the IW functions.
Due to heavy quark symmetry, there are many matrix elements that could
be used to obtain the same IW function.  By using different choices,
we can check to make sure the model is giving consistent results.
Also, many matrix elements are equal to zero at leading order in
$\Lambda_{\rm QCD}/m_{b,c}$ and $\alpha_s$, which is another way to
check the model results.  In the Appendix, we show the relevant matrix
elements for the different doublets and different spin structure.  
As an example, again consider
the $s_l^{\pi_l} = {\frac12}^-$ doublet.  Picking $\Gamma = 1$ we have
\begin{equation}
\xi(w) = \frac1{1+w} \langle D | \bar c b | \bar B \rangle 
= \frac1{1+w} \int \text{d}^3{\mathbf x}  \;
\Psi_{1,\;0,\;\frac12,\;0,\;0}^{\mathbf p}({\mathbf x}) \;
 {\bf 1} \; \Psi_{1,\;0,\;\frac12,\;0,\;0}({\mathbf x}),
\end{equation}
and similar relations can be found for other doublets.

\section{Results}

\subsection{The $D$ and $D^\ast$ multiplet}

\begin{figure}[t]
\centerline{\includegraphics[width=3.5in]{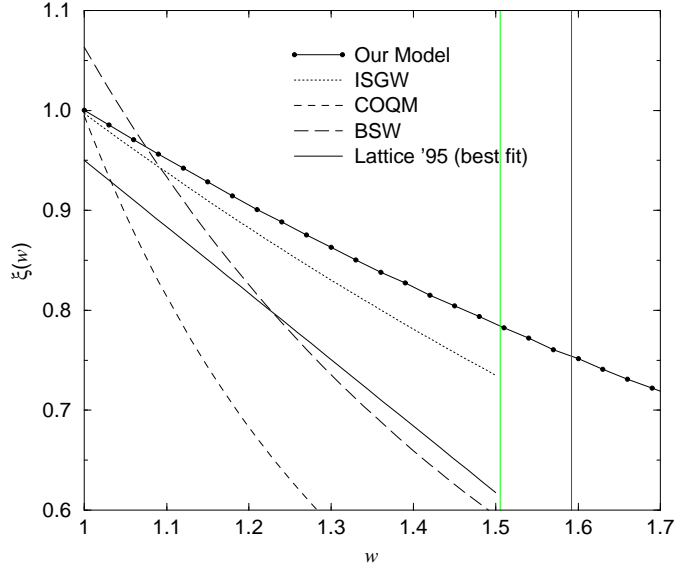}}
\caption{IW function for the $D-D^\ast$ multiplet, as predicted by different models. 
The vertical lines mark the end of the kinematical allowed region of the two mesons in the doublet.}
\label{DIW}
\end{figure}

The most investigated heavy-to-heavy decays are $\bar B\to D$ and 
$\bar B\to D^{*}$ which corresponds to the $s_l^{\pi_l} = {\frac12}^-$ 
doublet and are parametrized in terms of the same IW function $\xi(w)$:
\begin{eqnarray}
\frac{{\bf d}\Gamma(\bar B \rightarrow D \ell \bar \nu)}{{\bf d}w} &=&
\frac{G_F^2|V_{cb}|^2}{48\pi^3}(m_B+m_D)^2m_D^3(w-1)^{3/2} \xi^2(w),
\label{vub1}\\ 
\frac{{\bf d}\Gamma(\bar B \rightarrow D^\ast \ell \bar \nu)}{{\bf d}w} 
&=&
\frac{G_F^2|V_{cb}|^2}{48\pi^3}(m_B+m_{D^\ast})^2m_{D^\ast}^3\sqrt{w^2-1}(w+1)
\nonumber \\ 
&& \times \left( 1 + \frac{4w}{w+1} \frac{m_B^2-2w m_B m_{D^\ast} 
+ m_{D^\ast}^2}{(m_B+m_{D^\ast})^2} \right) \xi^2(w). 
\label{vub2}
\end{eqnarray}
We know from
heavy quark symmetry that, ignoring perturbative corrections, the IW
function for these decays, $\xi$, is normalized to one at zero recoil.
$|V_{cb}|$ can be determined from exclusive $\bar B\to
D^{(*)}\ell\bar\nu$ decay channels by a direct application of 
Eqs.~(\ref{vub1},\ref{vub2}).
On the experimental side it is difficult to  extrapolate $\xi(w)$ at zero
recoil, since the corresponding matrix element vanishes as 
$(w^2-1)^{1/2}$ for $D^*$ and $(w^2-1)^{3/2}$ for $D$.  
However, analyticity imposes stringent constraints \cite{anal,anal2}.

The normalization of $\xi$ at zero recoil is readily visible in 
Fig.~\ref{DIW}, where we plot $\xi$ as a function of $w \equiv v\cdot v'$.
If we parameterized the shape of $\xi$ as
\begin{equation}
\xi(1) \left[ 1-\rho^2(w-1) +c (w-1)^2+ \dots \right],
\end{equation}
we obtain for the slope parameter $\rho^2 = 0.501$
and for the curvature $c = 0.145$.\footnote{Note that our result is
not in agreement with Uraltsev's sum rule $\rho^2 > 3/4$
\cite{uraltsev}.}  In Fig.~\ref{DIW} the same IW function as predicted 
by the following alternative models: 
\begin{itemize}
\item ISGW~\cite{isgw}. In this model, nonrelativistic meson wave functions 
are obtained using a variational approach to the Schroedinger problem and
approximated with harmonic-oscillator wave functions.
\item SBW~\cite{sbw}. In this model, the form factors are calculated
assuming a pole structure.
\item Covariant Oscillator Quark Model or COQM~\cite{coqm}. This model is 
based on a covariant representation for nonrelativistic meson wavefunctions.
\end{itemize}
If Fig.~\ref{DIW} we also compare our model with the lattice
prediction of Ref.~\cite{jim}. The lattice result is affected by
unknown systematic quenching errors and discretization errors,
particularly for large momentum transferred. These errors are
difficult to estimate at present and are not reported in our
plot.\footnote{Modern computer technology allows for a noticeable
improvement over these lattice results.}

We also observe that a direct lattice determinations of the slope 
this IW exists~\cite{sachrajda}. This computation is done with a
propagating heavy quark slightly heavier than a charm meson and a light 
quark with a mass about the strange mass. It predicts a value of 
$\rho^2=1.7\pm0.02$, which is about a factor two larger than quark 
model predictions.

\subsection{The $D_1$ and $D_2^\ast$ multiplet}

\begin{figure}[t]
\centerline{\includegraphics[width=3.5in]{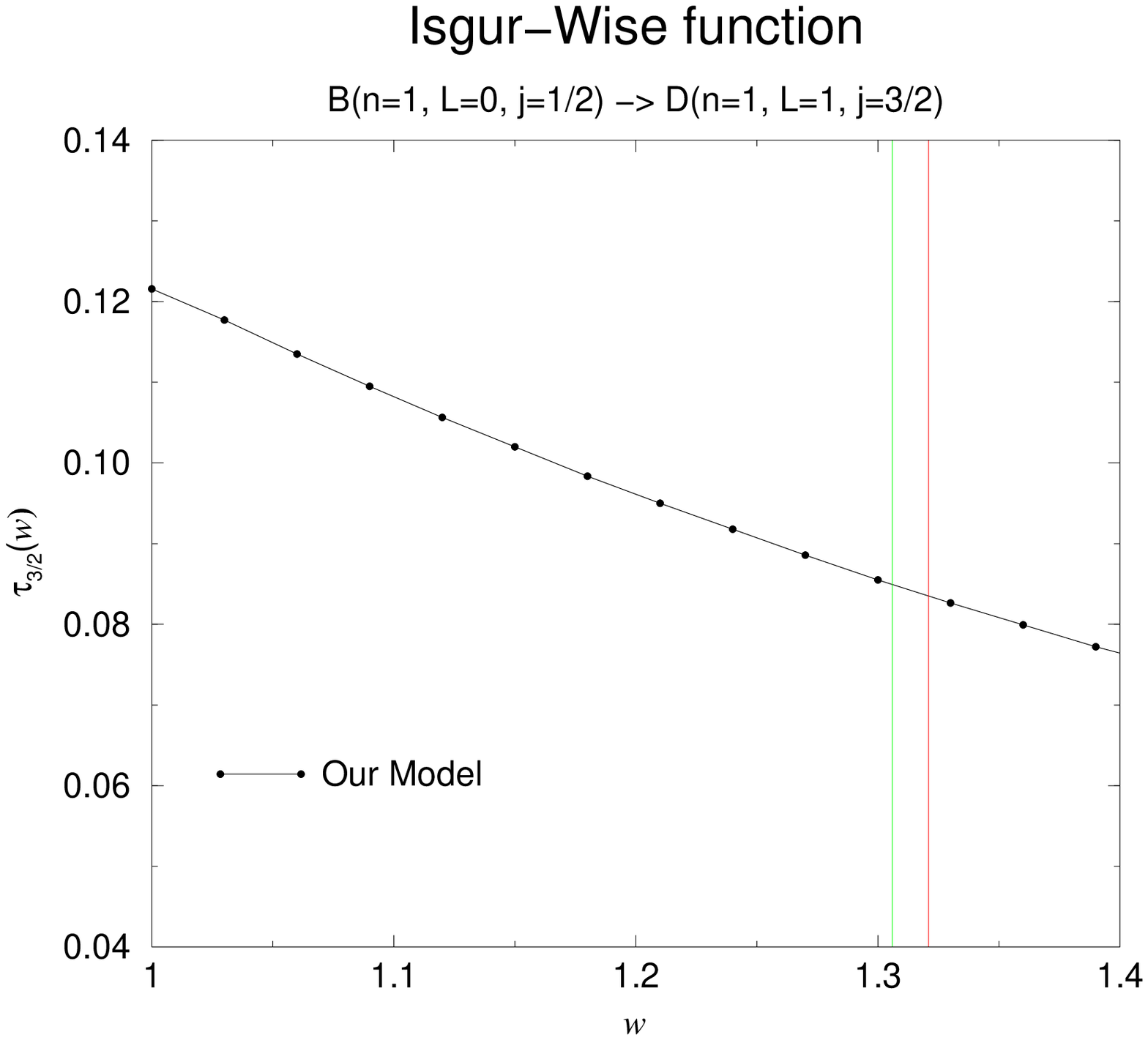}}
\caption{IW function for the $D_1-D_2^\ast$ multiplet.
The vertical lines mark the end of the kinematical allowed region of the two mesons in the doublet.}
\label{D1IW}
\end{figure}

The narrow resonances $D_1-D_2^\ast$ with light quantum numbers
$s_l^{\pi_l} = {\frac32}^+$ are important for a number of reasons.
For example, it is interesting to understand the composition of the
inclusive $B$ semileptonic decay rate in terms of exclusive final
states.  The particles in the ${\frac32}^+$ doublet are important
exclusive channels for this comparison.  It is also important to know the
decay spectrum for these particles as one the dominant backgrounds 
to $\bar B\to D^{(*)}$ decays.  Finally, there has been renewed effort in
constraining the slope parameter $\rho^2$ for $\bar B\to D^{(*)}$ using sum
rules and data on $B$ decays to excited $D$ states \cite{LeYaouanc,uraltsev}.

In Fig.~\ref{D1IW} we plot the IW function, $\tau_{3/2}$, for the
$D_1-D_2^\ast$ multiplet.  Unlike the previous case, there is no
reason why $\tau_{3/2}$ should be normalized to one at zero recoil.
If we parameterized the shape of $\tau_{3/2}$ as $\tau_{3/2}(w) =
\tau_{3/2}(1)[1 - \rho_{3/2}^{2}(w-1) + c_{3/2} (w-1)^2+ \dots]$, we
obtain for the normalization $\tau_{3/2}(1) = 0.122$, for the slope
parameter $\rho_{3/2}^{2} = 1.171$, and for the curvature $c_{3/2} =
0.601$.  In Table~\ref{tabtau} we compare our result with other
predictions found in the literature.  Note that our our value of
$\rho_{3/2}^{2}$ is consistent with the other models, while our
$\tau_{3/2}(1)$ is lower.

\begin{table}[t]
\begin {tabular}{l|cccc}
Ref & $\tau_{3/2}(1)$ & $\rho_{3/2}^{2}$ & 
      $\tau_{1/2}(1)$ & $\rho_{1/2}^{2}$ \\
\hline
Ours & 0.12 & 1.17 & 0.094 & 0.821 \\
\cite{ebert} & 0.49 & 1.53 & 0.28 & 1.04 \\
\cite{adam} & 0.41 & 1.5 & 0.41 & 1.0 \\
\cite{deandrea} & 0.56 & 2.3 & 0.09 & 1.1 \\
\cite{wambach} & 0.66 & 1.9 & 0.41 & 1.4 \\
\cite{colangelo} & & & 0.35 $\pm$ 0.08 & 2.5 $\pm$ 1.0 \\
\cite{morenas},\cite{godfrey} & 0.54 & 1.5 & 0.22 & 0.83 \\
\cite{morenas},\cite{colangelo2} & 0.52 & 1.45 & 0.06 & 0.73 \\
\cite{isgw} & 0.31 & 2.8 & 0.31 & 2.8 
\end{tabular} \vspace{6pt}
\caption{Comparison of IW functions $\tau_{3/2}$ and $\tau_{1/2}$ at
zero recoil and their respective slopes $\rho_{3/2}^2$ and
$\rho_{1/2}^2$ from different models.}
\label{tabtau}
\end{table}

\subsection{The $D_0^\ast$ and $D_1^\ast$ multiplet}

\begin{figure}[t]
\centerline{\includegraphics[width=3.5in]{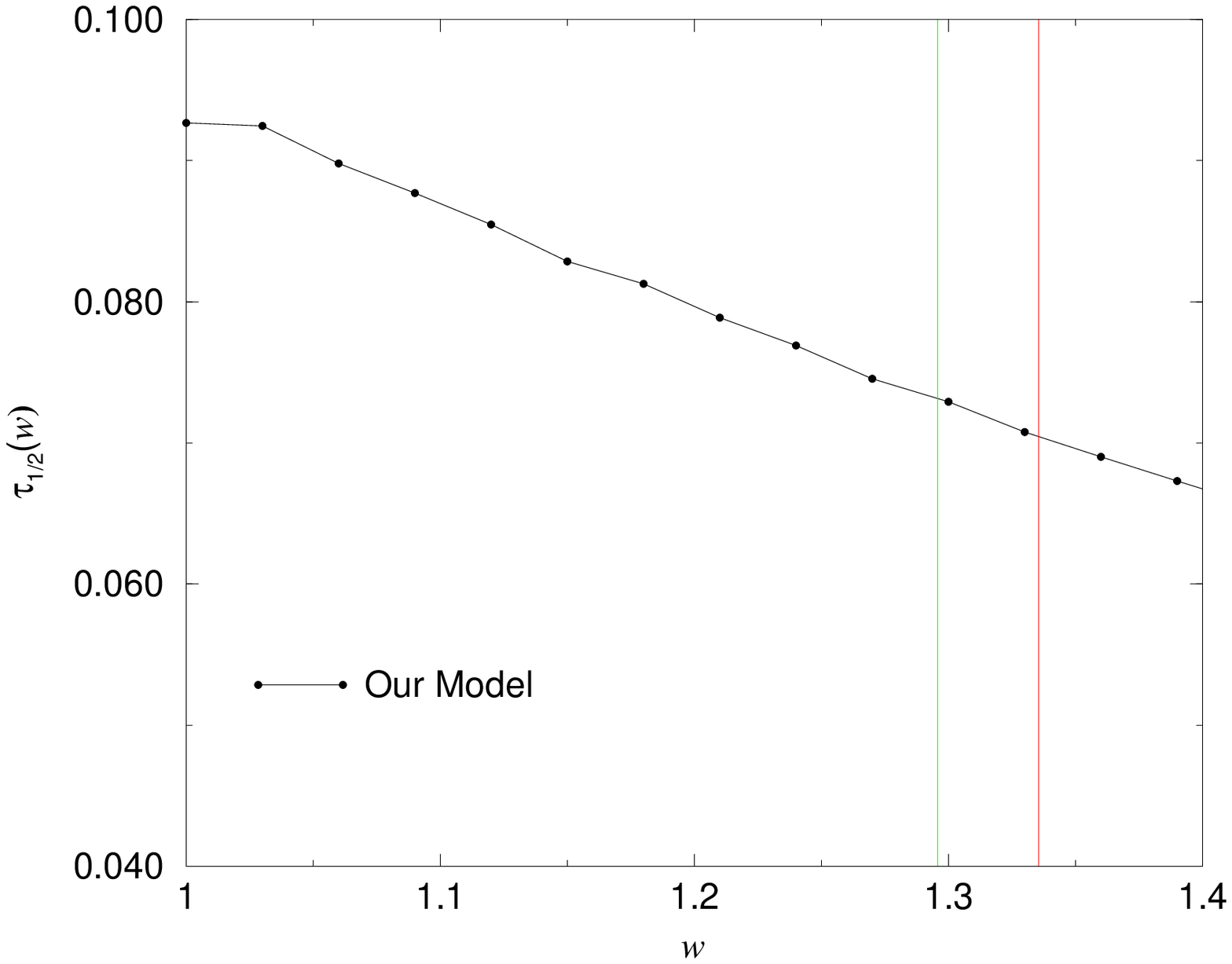}}
\caption{IW function for the $D_0^\ast-D_1^\ast$ multiplet.
The vertical lines mark the end of the kinematical allowed region of the two mesons in the doublet.}
\label{D0IW}
\end{figure}

Only recently data have been produced on the rates to these excited
mesons \cite{broad}.  The $s_l^{\pi_l} = {\frac12}^+$ multiplet is
very broad, as it can decay strongly to $D^{(*)}\pi$ in an $S$ wave~
\cite{dipierro} (the current experimental width 
is $290^{+101}_{-79}\pm 26 \pm 36$ keV for the $J=1$ meson),
while $D_1$ and $D_2^*$ can only decay through a $D$ wave, thus being
narrower resonances (the experimental width is $18.9^{+4.6}_{-3.5}$ keV for 
the $J=1$ meson) \cite{isgurwise2}.

In Fig.~\ref{D0IW} we plot the IW function, $\tau_{1/2}$, for
the $s_l^{\pi_l} = {\frac12}^+$, $D_0^\ast-D_1^\ast$ multiplet.
Again, there is no reason why $\tau_{1/2}$ should be normalized to one at
zero recoil.  If we parameterize the shape of $\tau_{1/2}$ as
$\tau_{1/2}(w) = \tau_{1/2}(1)[1 - \rho_{1/2}^2(w-1) + c_{1/2}
(w-1)^2+ \dots]$, we obtain for the normalization $\tau_{1/2}(1) =
0.094$, for the slope parameter $\rho_{1/2}^2 = 0.821$, and for the
curvature $c_{1/2} = 0.244$.  In Table~\ref{tabtau}, we compare our
result with other predictions found in the literature.

\subsection{The $D_1^{**}$ and $D_2^{**}$ multiplet}

\begin{figure}[t]
\centerline{\includegraphics[width=3.5in]{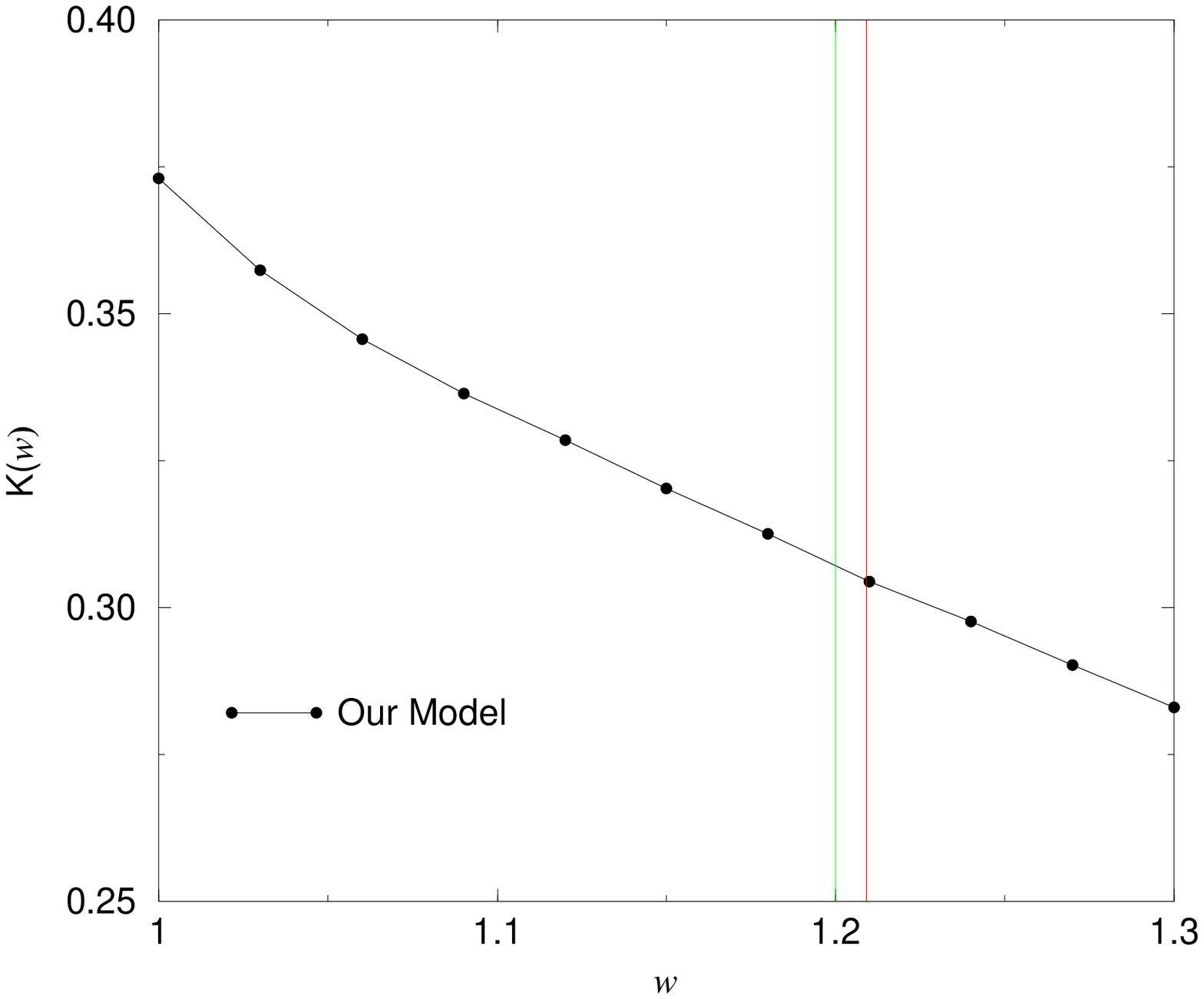}}
\caption{IW function for the $D_1^{**}-D_2^{**}$ multiplet.  The
vertical lines mark the end of the kinematical allowed region of the
two mesons in the doublet.}
\label{D2IW}
\end{figure}

To be complete, here we briefly mention the only other doublet
containing a spin one meson, the $s_l^{\pi_l} = {\frac32}^-$,
corresponding to $L=2$ orbital excitations in the quark model.  The
states are expected to be even broader than the $s_l^{\pi_l} = {\frac12}^+$ 
multiplet.  In Fig.~\ref{D2IW} we show the IW
function, $\kappa$ for this doublet.  
If we parameterize the shape of $\kappa(w)$ as
$\kappa(w) = \kappa(1)[1 - \rho^{-\,2}_{3/2}(w-1) + c^{-}_{3/2}
(w-1)^2+ \dots]$, we obtain for the normalization
$\kappa(1)=0.367$, for the slope $\rho^{-\,2}_{3/2}=0.884$, and for the
curvature $c^{-}_{3/2}=0.377$.

\subsection{The Radial Excitations}

While none have been seen, there are radial excitations of all the
above mentioned doublets.  It is unlikely that they will be seen in
the near future, but their effects are important in reconciling the
inclusive $b\to c$ and exclusive $\bar B \to X_c$ semileptonic
decay rates.  The radial excitations are also important as they enter
into sum rule calculations.  We therefore discuss the first radial
excitations of the above doublets here.  We denote the IW functions
for the radial excitations with primed versions of the same Greek
symbols as their non-radially excited counterparts.

\begin{figure}[t]
\begin{center}
\begin{tabular}{cc}
\includegraphics[width=3in]{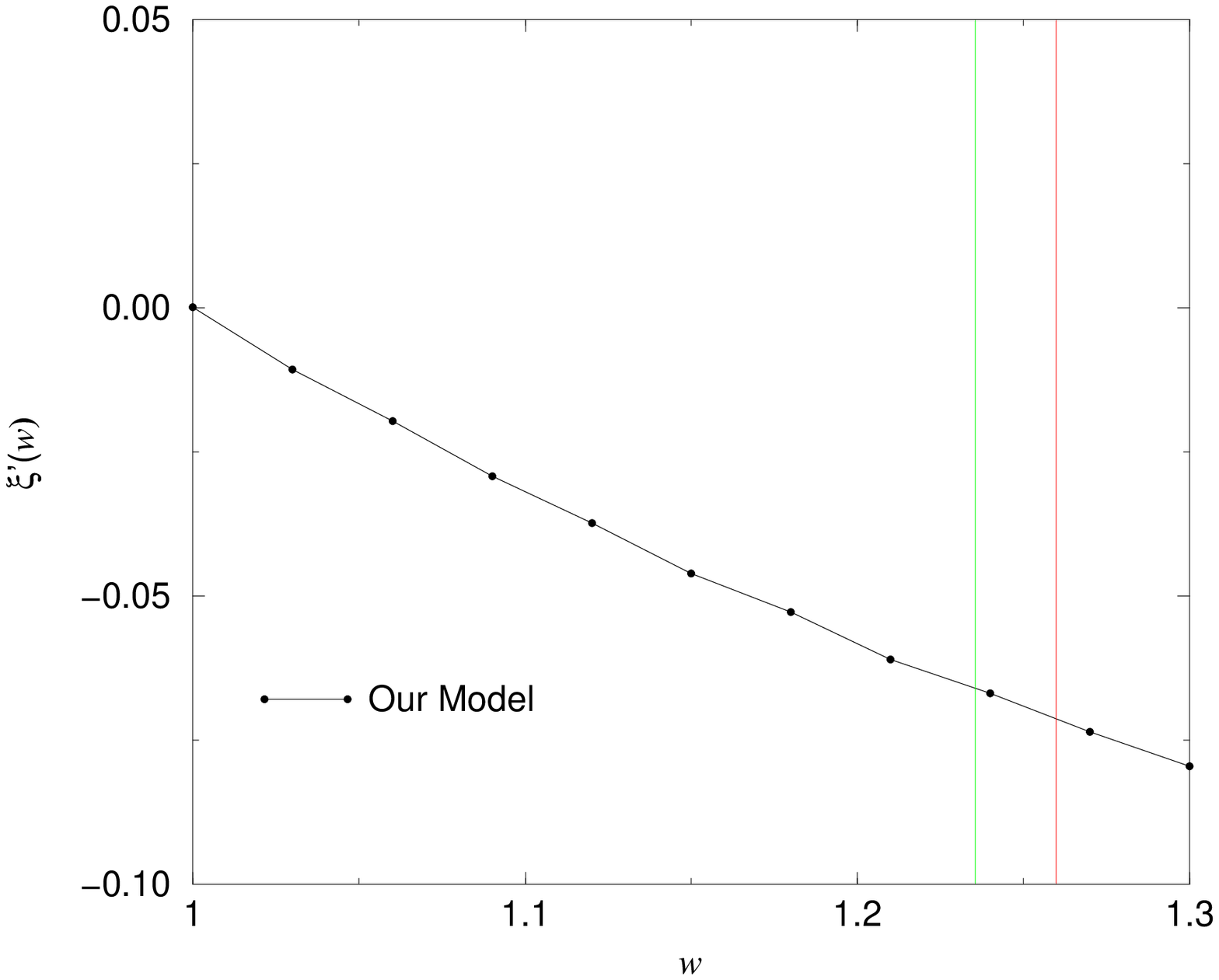} &
\includegraphics[width=3in]{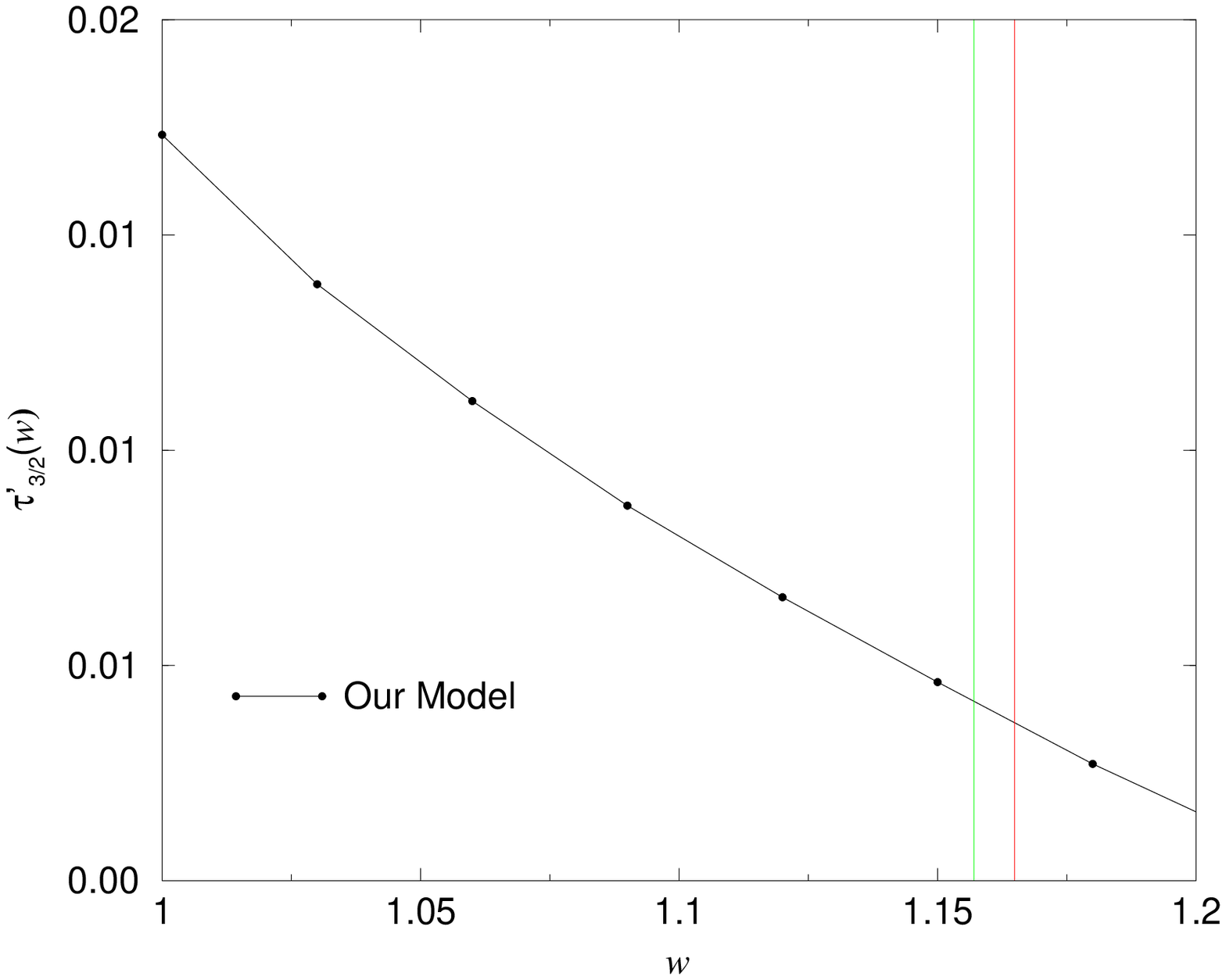} \\
\includegraphics[width=3in]{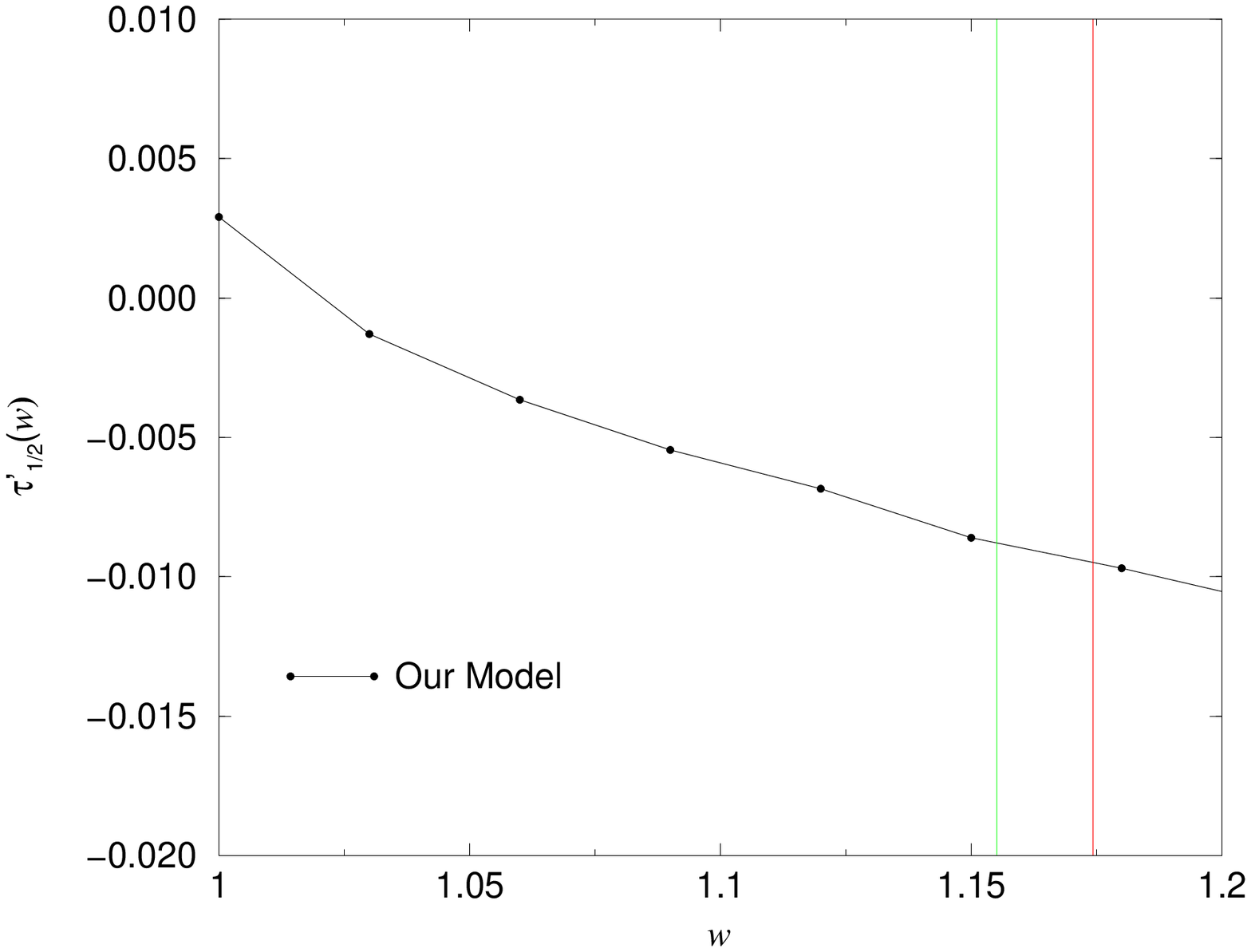} &
\includegraphics[width=3in]{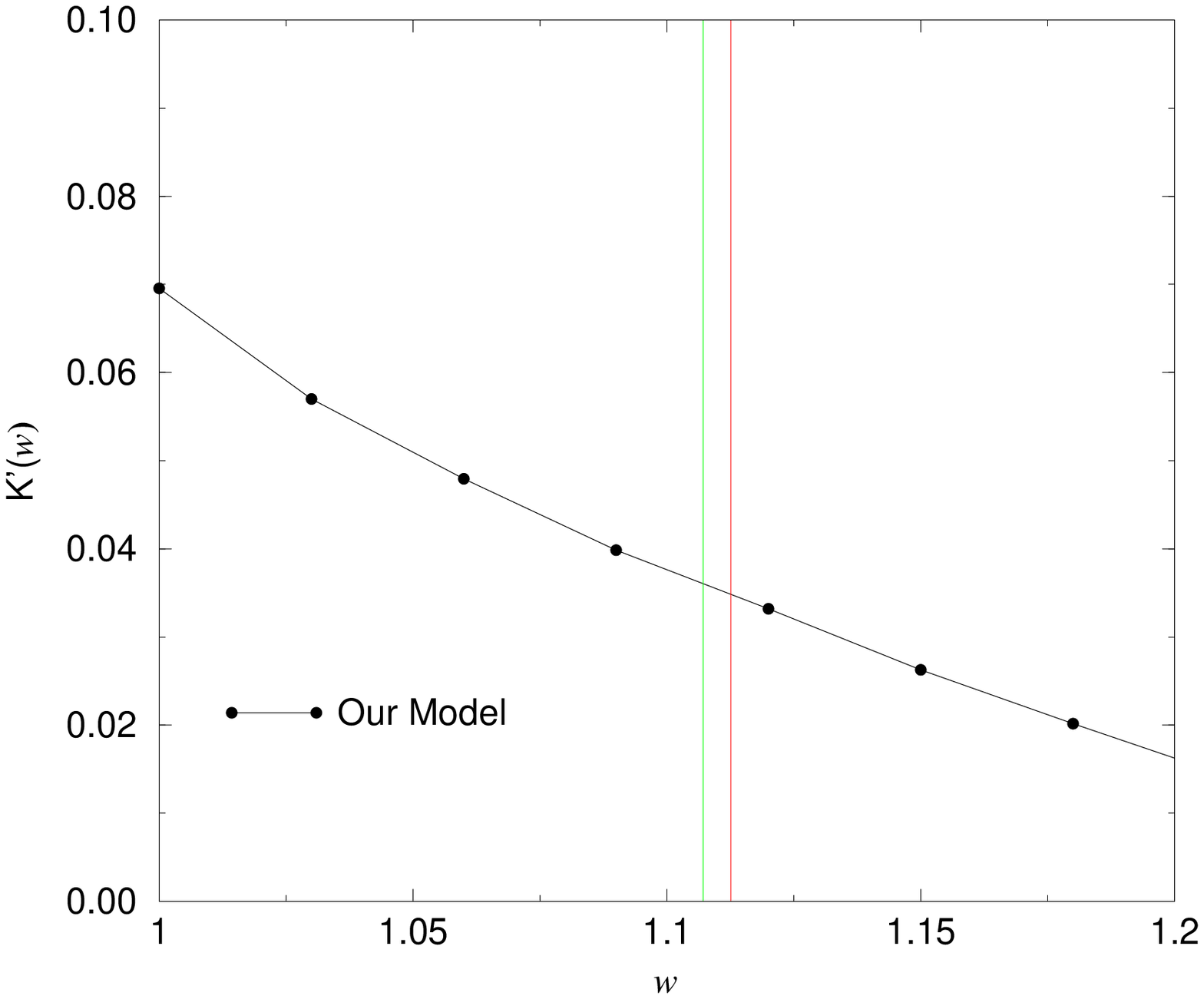} 
\end{tabular}
\end{center}
\caption{IW function for the first radial excitations with spin J<=2.
The vertical lines mark the end of the kinematical allowed region of the two mesons in each doublet.}
\label{D2all}
\end{figure}

At zero recoil, the IW function for $\bar B\to D^{(*)}$ was normalized,
$\xi(1)=1$.  For the radial excitation, there is no overlap between
the $B$ and $D^{'(*)}$ at zero recoil, thus $\xi'(1)=0$.  This can be
seen in Fig.~\ref{D2all}.  If we expand $\xi'$ around $w-1$, we get
\begin{equation}
\xi'(w) = -0.325 (w-1) + 0.213 (w-1)^2.
\end{equation}
Our result is very different from the conclusion of Ref.~\cite{ebert2}.
They get $\xi'(w) = 2.2 (w-1) + 2.6 (w-1)^2$.

For the radial excitations of the other doublets $s_l^{\pi_l} =
{\frac12}^+,\ {\frac32}^\pm$ we do not have any normalization
requirements at zero recoil (since the matrix elements vanish at leading order because of the heavy quark spin symmetry).  
In Fig.~\ref{D2all} we show the IW functions for $\xi'$, $\tau'_{3/2}$, 
$\tau'_{1/2}$ and $\kappa'$ respectively.

\section{Conclusion}

Information on the decays of $B$ mesons into charmed mesons is
important for a number of reasons.  Using Heavy Quark Effective
Theory, these decays can be written in terms of Isgur-Wise functions
which parameterized the form factors.  These functions cannot be
calculated except, eventually, on the lattice.

In this paper we derived the Isgur-Wise functions of $\bar B \to X_c$ within 
a quark model, where $X_c$ can be any spin 0 -- 2 charmed meson or 
one of their first radially excited states. Our results are compared with 
independent predictions found in the literature. Our model differs from the
others because we derived the wavefunctions of the $B$ meson and 
excited charmed  mesons in a relativistic fashion by fitting the experimental
spectrum with model predictions.

Our results for the Isgur-Wise function, $\xi$, for $B$ to $D^{(*)}$
decays is consistent with other model predictions but its slope is
milder. Our result is also consistent with the preliminary lattice
results of Ref.~\cite{jim}.  For decays into the $P$-waves the
situation is even more uncertain since different models predict a wide
varieties of results.  We believe that further model independent
investigations are required in order to put more constraints important
phenomenological quantities.

\acknowledgements

We want to thank Estia Eichten for useful discussions.  Fermilab is
operated by URA under DOE contract No.~DE-AC02-76CH03000.

\appendix
\section{Matrix elements in HQET}

In this Appendix we collect the relations between the matrix elements
of the form $\langle D |\bar c \Gamma d | \bar B \rangle$ to
expressions in terms of the Isgur-Wise functions in HQET at leading
order in $\Lambda_{\rm QCD}/m_{b,c}$ and $\alpha_s$, in analogy with
Eq.~(\ref{matrix1}). The matrices
containing the states for the different doublets are shown in
Eqs.~(\ref{hminus}) and (\ref{hplus}-\ref{fpm}).  The choices of Dirac
structure are $\Gamma = (1,\ \gamma_5,\ \gamma^\mu,\
\gamma^\mu\gamma_5)$. In this appendix $\epsilon^\mu$ is the polarization 
vector for spin 1 particles and $\epsilon^{\mu\nu}$ is the polarization 
tensor for spin 2 particles, while $\epsilon^{\alpha\beta\gamma\delta}$ is 
the usual Minkowskian antisymmetric tensor. $v$ is the velocity of the 
decaying $B$ meson and $v'$ is the velocity of the charmed decay product.
All matrix elements not explicitly shown are zero:
\begin{eqnarray}
\langle\; D(\Over12^{-})\;|\;\bar c b \;|\; \bar B \;\rangle  
  &=& (1+w)\xi(w) 
\\
\langle\; D(\Over12^{-})\;|\;\bar c \gamma^\mu b \;|\; \bar B \;\rangle  
  &=& (v+v')^\mu \xi(w) 
\\
\nonumber
\\
\langle\; D^\ast(\Over12^{-})\;|\;\bar c \gamma^5 b \;|\; \bar B \;\rangle  
  &=& -\epsilon^\mu v_\mu \xi(w) 
\\
\langle\; D^\ast(\Over12^{-})\;|\;\bar c \gamma^\mu b \;|\; \bar B \;\rangle  
&=& i \epsilon^{\mu\alpha\beta\gamma} 
  \epsilon_\alpha v_\beta v'_\gamma \xi(w) 
\\
\langle\; D^\ast(\Over12^{-})\;|\;\bar c \gamma^\mu \gamma^5 b \;|\; 
  \bar B \;\rangle &=& 
  [ \epsilon^\mu(1+w)-\epsilon^\nu v_\nu v'^\mu] \xi(w) 
\\
\nonumber
\\
\langle\; D_0(\Over12^{+})\;|\;\bar c \gamma^5 b \;|\; \bar B \;\rangle  
  &=& 2 (w-1) \tau_{1/2}(w) 
\\
\langle\; D_0(\Over12^{+})\;|\;\bar c \gamma^\mu \gamma^5 b \;|\; \bar B \;\rangle  
  &=& 2 (v-v')^\mu \tau_{1/2}(w) 
\\
\nonumber
\\
\langle\; D_1^\ast(\Over12^{+})\;|\;\bar c b \;|\; \bar B \;\rangle 
  &=& - 2 \epsilon^\nu v_\nu \tau_{1/2}(w) 
\\
\langle\; D_1^\ast(\Over12^{+})\;|\;\bar c \gamma^\mu b \;|\; \bar B \;\rangle  
  &=& 2 [\epsilon^\mu (w-1)-\epsilon^\nu v_\nu v'^\mu] \tau_{1/2}(w) 
\\
\langle\; D_1^\ast(\Over12^{+})\;|\;\bar c \gamma^\mu \gamma^5 b \;|\; 
  \bar B \;\rangle  &=& 
  2 i \epsilon^{\mu\alpha\beta\gamma}\epsilon_\alpha v_\beta 
  v'_\gamma \tau_{1/2}(w) 
\\
\nonumber
\\
\langle\; D_1(\Over32^{+})\;|\;\bar c b \;|\; \bar B \;\rangle  
  &=& -\sqrt{2} \epsilon^\nu v_\nu (1+w) \tau_{3/2}(w)
\\
\langle\; D_1(\Over32^{+})\;|\;\bar c \gamma^\mu b \;|\; \bar B \;\rangle  &=& 
  \sqrt{\frac12} [ \epsilon^\mu (1-w^2)-3\epsilon^\nu v _\nu v^\mu 
  + (w-2) \epsilon^\nu v _\nu v'^\mu\ \tau_{3/2}(w) 
\\
\langle\; D_1(\Over32^{+})\;|\;\bar c \gamma^\mu \gamma^5 b \;|\; \bar B \;\rangle
  &=& -\frac{i}{\sqrt{2}}(1+w)\epsilon^{\mu\alpha\beta\gamma}
  \epsilon_\alpha v_\beta v'_\gamma \tau_{3/2}(w) 
\\
\nonumber
\\
\langle\; D_2^\ast(\Over32^{+})\;|\;\bar c \gamma^5 b \;|\; \bar B \;\rangle  
  &=& \sqrt{3} \epsilon^{\alpha\beta}v_\alpha v_\beta \tau_{3/2}(w) 
\\
\langle\; D_2^\ast(\Over32^{+})\;|\;\bar c \gamma^\mu b \;|\; \bar B \;\rangle  
  &=& -i \sqrt{3} \epsilon^{\mu\alpha\beta\gamma} 
  \epsilon_{\alpha\sigma}v^\sigma v_\beta v'_\gamma \tau_{3/2}(w) 
\\
\langle\; D_2^\ast(\Over32^{+})\;|\;\bar c \gamma^\mu \gamma^5 b \;|\; 
  \bar B \;\rangle  &=& 
  v_\alpha [\epsilon^{\alpha \beta} v_\beta v'^\mu 
  -\sqrt{3} (1+w) \epsilon^{\alpha\mu} ] \tau_{3/2}(w) 
\\
\nonumber
\\
\langle\; D_1(\Over32^{-})\;|\;\bar c \gamma^5 b \;|\; \bar B \;\rangle  
  &=& \sqrt{\frac23} (1-w)\epsilon^\nu v_\nu \kappa(w) 
\\
\langle\; D_1(\Over32^{-})\;|\;\bar c \gamma^\mu b \;|\; \bar B \;\rangle  
  &=& \frac{i}{\sqrt{6}}(1-w)\epsilon^{\mu\alpha\beta\gamma}
  \epsilon_\alpha v_\beta v'_\gamma \kappa(w) 
\\
\langle\; D_1(\Over32^{-})\;|\;\bar c \gamma^\mu \gamma^5 b \;|\; 
  \bar B \;\rangle  &=& 
  \sqrt{\frac16} \epsilon^\nu[(2+w)v_\nu v'^\mu - 3 v_\nu v^\mu+ 
  (1-w^2)g_\nu^\mu ] \kappa(w) 
\\
\nonumber
\\
\langle\; D_2^\ast(\Over32^{-})\;|\;\bar c b \;|\; \bar B \;\rangle  
  &=& v_\alpha v_\beta \epsilon^{\alpha\beta} \kappa(w) 
\\
\langle\; D_2^\ast(\Over32^{-})\;|\;\bar c \gamma^\mu b \;|\; \bar B \;\rangle  
  &=& [ (1-w) \epsilon^{\alpha\mu} v_\alpha + \epsilon^{\alpha\beta}
  v'^\mu v_\alpha v_\beta ] \kappa(w) 
\\
\langle\; D_2^\ast(\Over32^{-})\;|\;\bar c \gamma^\mu \gamma^5 b \;|\; 
  \bar B \;\rangle &=& 
  i \epsilon^{\mu\alpha\beta\gamma}\epsilon_{\alpha\rho}
  v^\rho v_\beta v'_\gamma \kappa(w)
\end{eqnarray}

\end{document}